\documentclass[article,prd,nofootinbib,preprintnumbers,floats,aps]{revtex4}

\usepackage{graphicx}
\usepackage{graphics}
\usepackage{amsmath,amssymb}
\usepackage[usenames]{color}
\usepackage{subfigure}
\usepackage{hyperref}

\renewcommand{\Re}[1]{\hbox{Re} ~ #1}

\newcommand{\be}{\begin{equation}}
\newcommand{\ee}{\end{equation}}
\newcommand{\bea}{\begin{eqnarray}}
\newcommand{\eea}{\end{eqnarray}}
\newcommand{\ben}{\begin{enumerate}}
\newcommand{\een}{\end{enumerate}}
\newcommand{\bit}{\begin{itemize}}
\newcommand{\eit}{\end{itemize}}

\newcommand{\la}[1]{\label{#1}}

\newcommand{\Eq}[1]{Eq.~(\ref{#1})}

\newcommand{\Sec}[1]{Sec.~\ref{#1}}

\newcommand{\Fig}[1]{Fig.~\ref{#1}}

\newcommand{\vv}[1]{\mathbf #1}						

\newcommand{\bert}{\raise-0.45mm\hbox{\Large$\Box$}}		

\newcommand{\Eg}{E_{\rm gap}}

\definecolor{BrickRed}{cmyk}{0,0.89,0.94,0.28}				
\definecolor{MidnightBlue}{cmyk}{0.98,0.13,0,0.43}			
\definecolor{DarkGreen}{rgb}{0.100806,0.495968,0.209979}
\definecolor{orange}{rgb}{0.587167,0.354498,0.146197}

\begin{document}
\title{A thermodynamic cycle for the solar cell}

\author{Robert Alicki}\email{fizra@ug.edu.pl}
\affiliation{Institute of Theoretical Physics and Astrophysics, University of Gda\'nsk, 80-952 Gda\'nsk, Poland}

\author{David Gelbwaser-Klimovsky}\email{dgelbwaser@fas.harvard.edu}
\affiliation{Department of Chemistry and Chemical Biology, Harvard University, Cambridge, MA 02138, USA}

\author{Alejandro Jenkins}\email{alejandro.jenkins@ucr.ac.cr}
\affiliation{Escuela de F\'isica, Universidad de Costa Rica, 11501-2060, San Jos\'e, Costa Rica}
\affiliation{Academia Nacional de Ciencias, 1367-2050, San Jos\'e, Costa Rica \\}

\date{13 Jun.\ 2016; last revised 9 Jan.\ 2017.  To appear in {\it Annals of Physics}}

\begin{abstract}
A solar cell is a heat engine, but textbook treatments are not wholly satisfactory from a thermodynamic standpoint, since they present solar cells as directly converting the energy of light into electricity, and the current in the circuit as maintained by an electrostatic potential.  We propose a thermodynamic cycle in which the gas of electrons in the $p$ phase serves as the working substance.  The interface between the $p$ and $n$ phases acts as a self-oscillating piston that modulates the absorption of heat from the photons so that it may perform a net positive work during a complete cycle of its motion, in accordance with the laws of thermodynamics.  We draw a simple hydrodynamical analogy between this model and the ``putt-putt'' engine of toy boats, in which the interface between the water's liquid and gas phases serves as the piston.  We point out some testable consequences of this model. \\



\end{abstract}

\maketitle

\tableofcontents

\section{Introduction}
\la{sec:intro}

A solar cell, also called a photovoltaic cell, is a device that can convert the energy of light into an electrical current.  Although the earliest solar cells date from the 19th century \cite{Becquerel, solar-hist}, the theoretical account of their operating principles remains somewhat unsatisfactory.  This is not without precedent in the history of science: practical steam engines were built long before the formulation of the laws of thermodynamics \cite{thermo-hist}, while airplanes are much older than a satisfactory theory of lift on an aerofoil. \cite{aerofoil}

A solar cell is made of a semiconducting crystal (usually silicon) in two distinct phases: one with $n$-type and the other with $p$-type doping.  Figure \ref{fig:textbook} illustrates the explanation that textbooks commonly offer of how an illuminated solar cell generates voltage and current.  This picture is correct so far as it goes, but it does not explain the origin of the driving force that generates and maintains the direct current (DC) flowing in a closed circuit.  According to W\"urfel and W\"urfel,
\begin{quote}
We frequently read that it is just the electric field of a $pn$-junction which supplies the driving force for the currents flowing during illumination [\ldots] In fact something must be wrong in our physical education, if we think that a DC current can at all be driven in a closed circuit by a purely electrical potential difference.  The word potential alone should tell us that no energy can be gained by moving a charge along any closed path. \cite{Wuerfel}
\end{quote}
The authors go on to argue that the force driving the current in a circuit connected to a solar cell is given by a chemical potential, like that in a battery.\footnote{On electrochemical potentials and their relation with the electromotive force in the context of thermoelectric generators, see \cite{HeikesUre, Apertet}.}  This argument cannot be complete either, because the same objection to an electrostatic potential driving the cyclic DC applies to a static chemical potential.  Note that a battery can be recharged, but only by externally reversing the current (see, e.g., \cite{Fermi, Purcell}).

As we shall review in \Sec{sec:battery}, there is a fundamental difference between a battery, which while it generates current is relaxing to a new equilibrium state with vanishing voltage, and a solar cell, which is a heat engine capable of steadily supplying voltage and current as long as it remains in contact with two heat baths at different temperatures.  This calls for the formulation of a thermodynamic cycle in which the working substance periodically returns to its initial state after performing net work on its surroundings.

That the operation of a solar cell should be characterized by a cycle was argued in \cite{cellcycle}, but without arriving at a concrete dynamical implementation. The idea of treating solar cells, thermoelectric generators, and fuel cells as engines in which a plasma oscillation serves as a piston has been advanced recently, within the formalism of the Markovian master equation for open quantum systems, in \cite{Markovian1, Markovian2, fuelcells}.\footnote{After this work had been completed and submitted to the physics pre-print archive, there appeared on the same archive an independent argument for the necessity of incorporating a feedback-induced periodicity into the characterization of heat engines, with a particular focus on thermoelectric systems. \cite{Goupil}}

Here we offer, for the first time, a dynamically complete model of the operation of the solar cell as an autonomous heat engine, based on an understanding of the mutual coupling of the relevant subsystems and of the corresponding feedback.  Moreover, this model is presented in terms accessible to any student familiar with basic thermodynamics, in part by exploiting the close analogy to the ``putt-putt'' engine of model boats that were once popular children's toys \cite{Piot}.  The implementation of this model based on the plasma oscillation at the solar cell's $p$-$n$ junction makes concrete predictions that could be investigated experimentally using sub-millimeter radiation.

\begin{figure} [t]
\begin{center}
	\includegraphics[width=0.45 \textwidth]{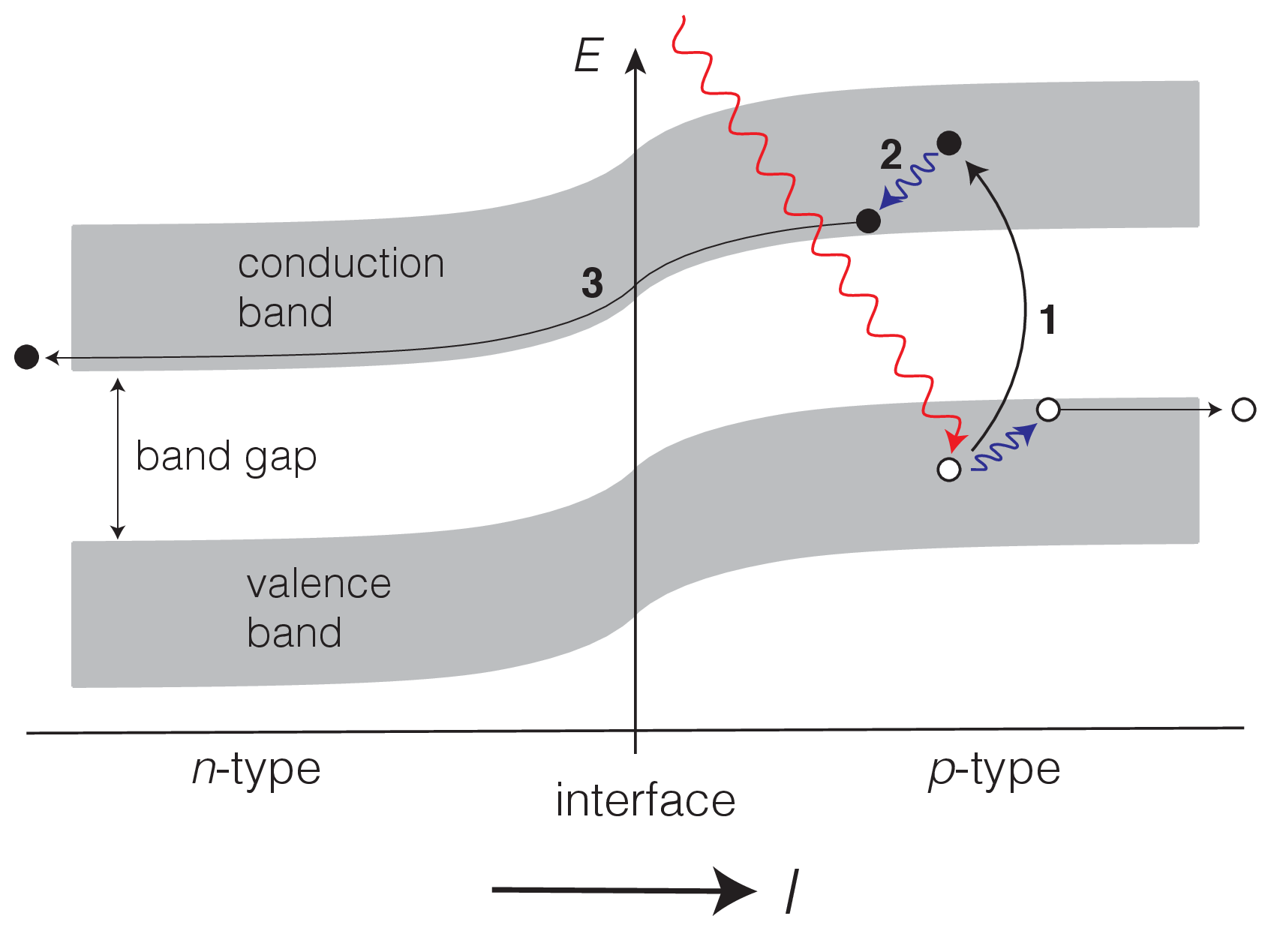}
\end{center}
\caption{\small Processes in the standard textbook account of the photovoltaic effect: {\bf 1.}  A photon is absorbed in the $p$-type phase of the semiconductor, generating a conducting pair (electron and hole).  {\bf 2.} The pair quickly thermalizes with the phonons in the lattice, dissipating the energy excess above the band gap. {\bf 3.} The electron is driven to the left by the potential difference across the interface, while the hole moves to the right, generating a voltage between the two terminals.\la{fig:textbook}}
\end{figure}

\section{Batteries vs.\ engines}
\la{sec:battery}

Let us first review the important distinction between non-cyclic sources of power ---such as batteries--- and heat engines.  This will help to clarify both the need to describe the photovoltaic effect as a cycle and how to do so consistently with the laws of thermodynamics.

\subsection{Non-cyclic power sources}
\la{ssec:noncyclic}

As shown in \Fig{fig:discharging}(a), in a charged capacitor the electrostatic field $\vv E$ can drive an electron (charge $-e$) from the negatively charged plate $A$ to the positively charged plate $B$, performing work
\be
W = eV = e \int_B^A d \vv s \cdot \vv E ~,
\la{eq:capacitor}
\ee
which may drive a load or be dissipated by resistance.  The potential $V$ can perform a net $W > 0$ because the electrons do not follow a closed path and $V$ decreases with time while the circuit is closed.

A charged battery in an open circuit is in a thermodynamic equilibrium with a $V>0$ between the two plates, generated by a chemical potential difference.  When the circuit is closed, the electrons that emerge from the negatively charged plate are supplied by a chemical reaction, which in the common lead-acid battery ---represented schematically in \Fig{fig:discharging}(b)--- is
\be
\hbox{Pb} + \hbox{HSO}_4^- \rightarrow \hbox{PbSO}_4 + \hbox{H}^+ + 2e^- ~.
\la{eq:etocircuit}
\ee
The electrons that flow into the positively charged plate are absorbed by a different reaction:
\be
\hbox{PbO}_2 + \hbox{HSO}_4^- + 3 \hbox{H}^+ + 2e^- \rightarrow \hbox{PbSO}_4 + 2\hbox{H}_2\hbox{O} ~.
\la{eq:efromcircuit}
\ee
Within the battery, the current $I$ is given by a flow of ions (rather than electrons), driven against the electrostatic field by the chemical potential.  Reaction \eqref{eq:etocircuit} injects $H^+$ ions, and reaction \eqref{eq:efromcircuit} consumes them, generating a gradient of concentration that causes the ions to diffuse from the negatively charged to the positively charged plate. The chemical potential difference is being gradually exhausted as the discharging battery approaches an equilibrium with vanishing $V$. \cite{Purcell,chemicalpotential}

\begin{figure} [t]
\begin{center}
	\subfigure[]{\includegraphics[width=0.35 \textwidth]{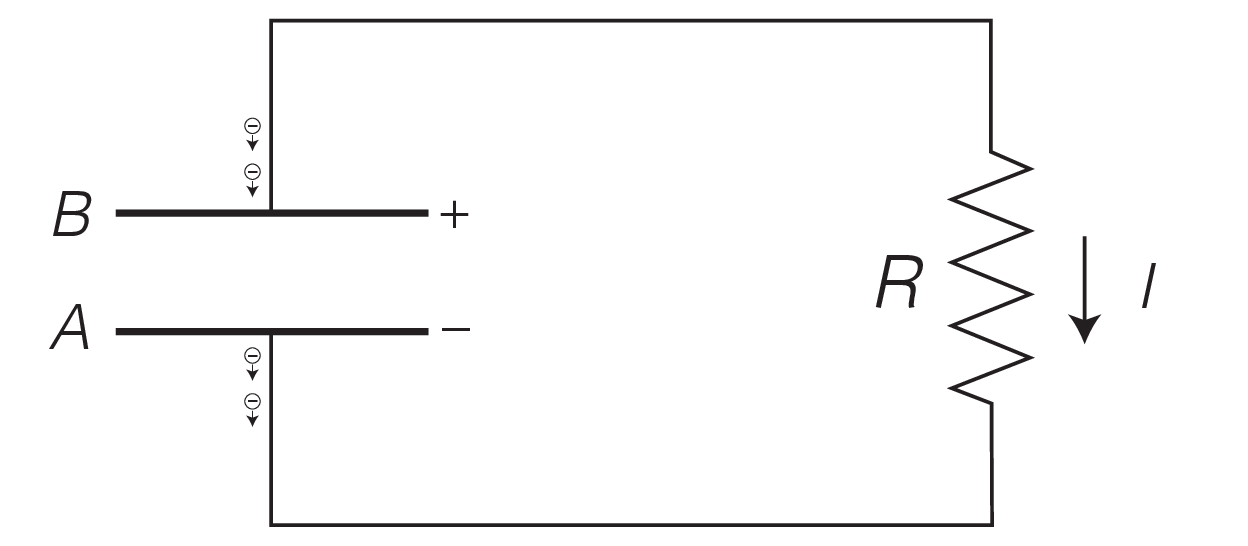}} \hskip 1 cm
	\subfigure[]{\includegraphics[width=0.52 \textwidth]{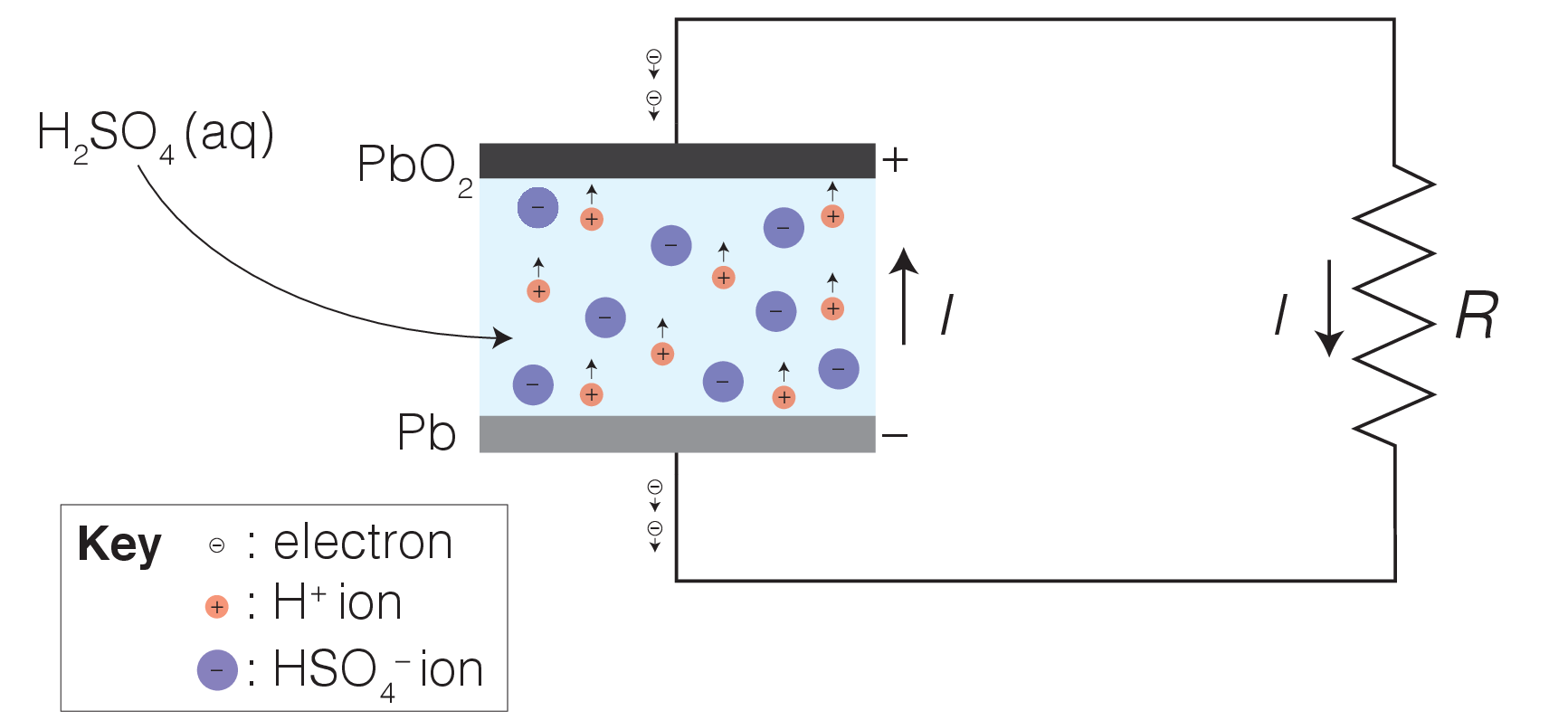}}
\end{center}
\caption{\small (a) The electrostatic potential $V$ between a negatively charged plate $A$ and positively charged plate $B$ drives a steadily decreasing current $I$ through the resistor $R$.  (b)  In a car battery, the electrons coming from the negatively charged plate are produced by the chemical reaction between spongy lead (Pb), attached to the plate, and the ions in an aqueous solution of sulfuric acid (H$_2$SO$_4$ (aq)).  Electrons are consumed by a reaction between the ions in the solution and the granules of lead dioxide (PbO$_2$) in contact with the positively charged plate.  The path of the current $I$ is closed within the battery by $H^+$ ions diffusing against the electric field.  Illustration adapted from Fig.\ 4.20 in \cite{Purcell}.\la{fig:discharging}}
\end{figure}

\subsection{Need for photovoltaic cycle}
\la{ssec:photocycle}

Illuminated solar cells (like thermoelectric generators and fuel cells) can maintain a DC without the local charge density being exhausted.  Such a current cannot be driven by a static electrochemical potential.  Moreover, it is well known that photons impart negligible momentum to electrons in a solar cell \cite{Wuerfel}.  Unlike the motion of the belt in a Van de Graaf generator (see \cite{Purcell}), the incident light cannot, therefore, power a DC by pushing charges against the electric field.

A solar cell that drives a steady DC must therefore be understood as a heat engine, operating between a hot bath of photons and a cold bath of phonons in the crystal of the cell's semiconductor.  As it stands, the mechanism shown in \Fig{fig:textbook} appears disturbingly like a {\it perpetuum mobile} of the second kind (see, e.g., \cite{Fermi}), with the energy of the photons directly converted into the work that drives the DC in the closed circuit.  The recombination of a conducting electron from the external circuit with a hole at the top of the cell's valence band corresponds to consumption of the illuminated cell's work output, not to the essential heat rejection into the cold bath demanded by the second law of thermodynamics.  The only heat rejection shown (process {\bf 2} in \Fig{fig:textbook}) is an inessential dissipation, not controlled by the cold bath temperature in the way required by Carnot's theorem.

Shockley and Queisser derived an efficiency bound for solar cells by requiring that the photodiode radiate as a black body at room temperature, while maintaining a detailed balance of the local charge densities \cite{SQ}.  That radiation is what keeps their model of the solar cell from being an unphysical {\it perpetuum mobile}.  Here we propose a more physically transparent way of arriving at a similar efficiency bound: to characterize the operation of a solar cell as a thermodynamic cycle (i.e., as a heat engine), in which the working substance (the electron gas in the $p$-type phase of the semiconductor) absorbs heat from the photons at a high effective temperature (see \Sec{ssec:local}) and rejects heat into its environment at room temperature, converting the difference into work that can drive a DC in an external circuit.  This picture calls for a self-oscillatory dynamic at the photodiode junction that has yet to be observed directly.  The resulting expression for the solar cell's limiting efficiency is presented in \Sec{sec:efficiency}.

\section{Self-oscillation in heat engines}
\la{sec:self-oscillation}

Self-oscillation is the generation and maintenance of a regular periodic motion at the expense of a source of power without any corresponding periodicity.  This definition is due to Andronov and his collaborators \cite{Andronov}, but the concept is much older and goes also by many other names.  This is qualitatively different from resonant phenomena, which depend on matching an external periodicity to the natural period of the oscillation.  Self-oscillation relies on a {\it feedback} mechanism, by which the oscillation modulates the action upon it of the external power source in such a way the oscillator extracts a net positive energy over a complete period of its motion.  For a detailed review of this subject from a physical perspective, see \cite{SO}.

Mechanical engineers are well aware that engines require a complicated dynamics, based on pistons, turbines, fly-wheels, valves, etc., in order automatically to convert a steady power source (such as a temperature difference) into a periodic mechanical action.  But, even though heat engines were at the heart of the classical thermodynamics of the 19th century (hence the importance in that theory of the concept of {\it cycle}), in modern statistical physics the principles behind the autonomous operation of an engine have remained something of a theoretical blind spot.  This is a particularly pressing issue in the case of microscopic heat engines ---such as the solar cell--- in which the self-oscillatory dynamic is difficult to observe directly.

\subsection{Mechanics and thermodynamics of self-oscillators}
\la{ssec:mech-thermo}

According to the laws of classical mechanics, if a mass element describes a cycle $C$ with period $\tau$, the work exerted on it by an external force $\vv F$ is
\be
W = \oint_C d \vv s \cdot \vv F = \int_0^\tau dt \, (\vv v \cdot \vv F) ~.
\la{eq:work-classical}
\ee
From the first equality of \Eq{eq:work-classical}, it follows that if the particle moves in a potential $\phi$, so that $\vv F = - \nabla \phi$, then $W = 0$.  From the second equality in \Eq{eq:work-classical}, we see that for the cyclic motion to be encouraged $\vv F$ must be properly modulated, the best case being that in which it leads the position by a quarter of a period, so that $\vv F$ varies in phase with the oscillator's velocity $\vv v$.  In a forced resonance, this is achieved only when the external driving frequency $\omega_d$ happens to be very close to the oscillator's natural frequency $\omega_0$ \cite{Georgi}.  But self-oscillators require no such tuning, because the oscillation itself induces the required modulation of $\vv F$.

A heat engine contains a working substance that undergoes a cyclic sequence of transformations such that, after a full cycle, it has absorbed more heat from a bath at a higher temperature than the heat that it has rejected into a bath at a lower temperature.  The net energy gained drives the macroscopic oscillation ---with the same period as the working substance's thermodynamic cycle--- of what we shall generically call the {\it piston} (though this might not always correspond to what an engineer would ordinarily think of as a piston).

Though this has not usually been stressed in the scientific literature, the piston's self-oscillation is an essential part of the operation of heat engines that run automatically, without an external agent performing the modulation.  It is the self-oscillatory dynamic that allows an engine to convert a non-periodic source of power (such as heat, which simply flows from high to low temperatures) into work outputted at a well-defined frequency.  Moreover, it is the need to dynamically {\it generate} a finite period for the piston's motion that introduces unavoidable energy losses in the operation of heat engines and other motors. \cite{LeCorbeiller}

Many motors lack resonant elements, so that their frequency of operation depends non-linearly on the power.  (Motors in which care is taken to have the frequency of operation be very stable and close to a linear resonance are usually intended as clocks.)  A physically oriented theory of self-oscillation may therefore provide a useful perspective on thermodynamic problems. \cite{SO}

\subsection{Rayleigh-Eddington criterion}
\la{ssec:Rayleigh}

For an infinitesimal transformation of a working substance, let $\delta W$ be the mechanical work performed by the substance on its surroundings, $\delta Q$ the heat absorbed by the substance, and $dN$ the change in the quantity of matter in the substance.  By the first law of thermodynamics, the change in the substance's internal energy is
\be
dU = \delta Q - \delta W + \mu dN ~,
\la{eq:1stlaw}
\ee
where $\mu$ is the chemical potential.  Over a complete thermodynamic cycle the substance returns to its initial state, so that the net change to the internal energy $U$ must be
\be
\Delta U = \oint \left( \delta Q - \delta W + \mu dN \right) = 0 ~.
\la{eq:conservation}
\ee
The net work done by the substance is therefore
\be
W = \oint \delta W = \oint ( \delta Q + \mu dN) ~.
\la{eq:work}
\ee
By the second law of thermodynamics (expressed in the form of Clausius's theorem) the entropy generated by the cycle is
\be
\sigma = - \oint \frac{\delta Q}{T} = - \oint \frac{\delta Q}{\bar T \left ( 1 + T_d / \bar T \right)} \geq 0 ~,
\la{eq:Clausius}
\ee
where $T$ is the substance's instantaneous temperature, with $\bar T$ its mean value over the cycle's period and \hbox{$T_d \equiv T - \bar T$}.  Combining Eqs.\ (\ref{eq:work}) and (\ref{eq:Clausius}) we obtain
\be
W \leq \oint \delta Q \left( 1 - \frac{1}{1 + T_d / \bar T} \right) + \oint dN \cdot \mu ~.
\la{eq:Eddington-mu}
\ee
This bound on $W$ is saturated in the absence of dissipation ($\sigma = 0$).  This is easily generalized to inhomogenous temperatures and chemical potentials by integrating over the maximum work that each part of the working substance may perform.

For an engine (such as a solar cell) that runs entirely on heat, either $dN = 0$ or $\mu = $ const., so that \Eq{eq:Eddington-mu} reduces to
\be
W \leq \oint \frac{\delta Q \cdot T_d}{\bar T + T_d} \simeq \frac{1}{\bar T} \oint \delta Q \cdot T_d~,
\la{eq:Eddington}
\ee
where the two integrals are approximately equal for $| T_d / \bar T | \ll 1$.  Eddington wrote \Eq{eq:Eddington} in the context of the self-oscillation of Cepheid variable stars \cite{Eddington1,Eddington2}.  It implies that mechanical oscillations are encouraged as long as heat is injected ($\delta Q > 0$) when the temperature of the working substance is higher ($T_d > 0$), and heat is rejected ($\delta Q < 0$) when the temperature is lower ($T_d < 0$).

Before Eddington, Rayleigh had formulated a similar criterion for thermoacoustic phenomena, according to which the mechanical oscillation of a volume of gas is most encouraged when the rate of heat flowing out of it varies in phase with the volume \cite{Rayleigh1, Rayleigh2}.  For any substance with positive pressure and positive heat capacity, adiabatic expansion reduces its temperature while adiabatic compression increases it, so that Rayleigh's criterion follows directly from \Eq{eq:Eddington}.

\begin{figure} [t]
\begin{center}
	\includegraphics[width=0.4 \textwidth]{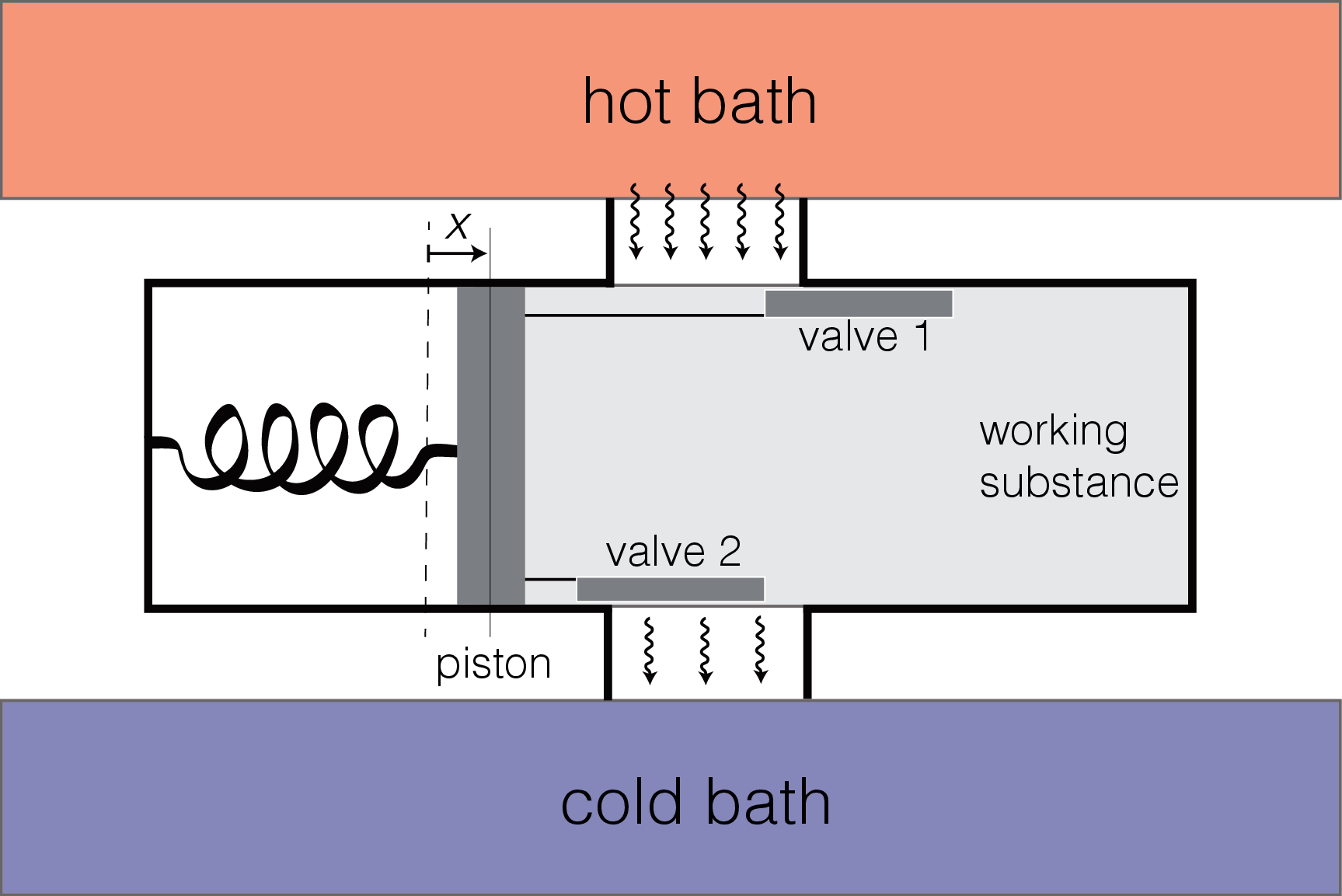}
\end{center}
\caption{\small The piston's self-oscillation, and therefore the automatic operation of this heat engine, depends on the valves modulating the rate of heat flow between the working substance and the two baths in accordance with \Eq{eq:Eddington}, so that $W > 0$.  This corresponds to a positive feedback between the oscillation of $x$ and the modulation of the heat flow.\la{fig:Rayleigh}}
\end{figure}

According to Rayleigh,
\begin{quote}
In any problem which may present itself of the maintenance of a vibration by heat, the principal question to be considered is the {\it phase} of the communication of heat relatively to that of the vibration. \cite{Rayleigh2}
\end{quote}
Note that any mechanical cycle may be regarded as a ``vibration''.  In fact, a major technical advance in the design of internal combustion engines, originally proposed in 1862 by Beau de Rochas based on a mechanical (rather than thermodynamic) argument, was to time the fuel's ignition to coincide with the moment of maximum compression of the working gas inside the cylinder \cite{Rochas}.  Modern gasoline, diesel, and other engines work on this principle. \cite{motors}

Let $x$ be the displacement of the piston, with increasing $x$ corresponding to compression of the working substance, as shown in \Fig{fig:Rayleigh}.  For the engine to run ($W > 0$), the relative phase $\varphi$ between $\delta Q$ and $x$ (or, equivalently, between $\delta Q$ and $T_d$ in \Eq{eq:Eddington}) must satisfy
\be
- \frac{\pi}{2} < \varphi < \frac{\pi}{2} ~.
\la{eq:phase}
\ee
We call this the ``Rayleigh-Eddington criterion''.  When heat flow is modulated by the piston's motion in accordance with this criterion, a {\it positive feedback} is established, which, if it overcomes the damping, causes the piston to self-oscillate.  Heat is most efficiently converted into mechanical energy when $\varphi = 0$.  This is achieved by arranging for valve 1 to be fully open and valve 2 fully closed when $x$ reaches its maximum.  Conversely, valve 1 should be fully closed and valve 2 fully open when $x$ reaches its minimum.

Reversing the signs of $\delta Q$ and $W$ in \Eq{eq:Eddington} shows that a cycle that violates the Rayleigh-Eddington criterion can run as a refrigerator, extracting heat from the working substance at the expense of external mechanical power.  The sign of the frequency of the piston's oscillation in \Fig{fig:Rayleigh} is not physically meaningful, and running it as a refrigerator would require interchanging the configurations of valves 1 and 2.  In some implementations of a heat cycle, such as Stirling engines, reversal between engine and refrigerator can be accomplished by reversing the direction in which the flywheel turns \cite{Stirling}.  In that case the sign of the flywheel's angular velocity is given physical significance by the phase relation between the motion of the power piston and the rate of heat injection to the working substance.

Note that, according to \Eq{eq:Eddington-mu}, a cycle of an engine that operates between baths at different values of $\mu$ may do positive work at constant $T = \bar T$, as long as the working substance's $\mu$ is modulated so that
\be
\oint dN \cdot \mu > 0 ~.
\la{eq:mu-modulation}
\ee
This is not pertinent to the solar cell, but it must be taken into account when conceptualizing fuel cells as self-oscillating chemical engines; see \cite{fuelcells}.

\subsection{Pistons in quantum thermodynamics}
\la{ssec:quantum}

The theoretical work of the last forty years on quantum thermodynamics has conceptualized a heat engine as an open system that absorbs heat in one state of the working substance and rejects it in a different state, thus permitting net work extraction over a complete cycle.  Both the modulation of the state and the work extraction are performed by a semi-classical, oscillatory degree of freedom (the ``piston''), which can be described by a time-dependent Hamiltonian \cite{piston1, piston2, piston3, piston4, piston5, piston6}.  (For a model of a heat engine with a time-independent Hamiltonian and a quantized piston, see \cite{quantumpiston}.)  Such a piston, which in an engine that operates without external modulation must be self-oscillatory, is conspicuously missing from the standard account of the photovoltaic effect, illustrated in \Fig{fig:textbook}.

In the mathematical formalism of quantum thermodynamics, the piston's periodicity is needed to define the engine's power output.  Alternative theoretical approaches based on the Onsager reciprocal relations \cite{Onsager} cannot account for the maintenance of a closed circulation ---such as the DC driven by an illuminated solar cell--- because those relations describe currents as gradients of scalar potentials.  Moreover, a heat engine remains arbitrarily far from equilibrium as long as the temperature difference between the baths persists, whereas the Onsager relations describe a system close to equilibrium and steadily relaxing towards it.

\subsection{Piston-less engines}
\la{ssec:self-rotation}

Consider the rubber-band engine shown in \Fig{fig:rubber-band}, taken from the {\it Feynman Lectures}, in which the spokes of a bicycle wheel have been replaced by rubber bands \cite{Feynman}.  Electric lamps heat the rubber bands on the left half of the wheel, causing them to pull more on the axle than the colder bands on the right half.  This displaces the wheel's center of mass with respect to the axle, causing a persistent gravitational torque.  In this case the rubber bands are the working substance.  The wheel's turning is both the engine's work output and the mechanism that modulates the $T_d$ of \Eq{eq:Eddington} for the rubber bands.

This engine may seem piston-less because the two functions attributed to the piston in \Sec{ssec:quantum} are performed by the macroscopic circulation of the working substance, a circulation that takes place in an external temperature gradient.  Another familiar instance of this is a hurricane, in which the vertical component of the air's circulation causes it to undergo what is essentially a Carnot cycle \cite{hurricane}.\footnote{Emanuel pointed out a logical flaw in previous physical models of hurricanes that is similar to the one that we identify in the textbook account of solar cells: ``Attempts to regard the condensation heat source as external lead to the oft-repeated statement that hurricanes are driven by condensation of water vapor, a view rather analogous to that of an engineer who proclaims that elevators are driven upward by the downward acceleration of counterweights. Such a view, though energetically correct, is conceptually awkward; it is far more natural to consider the elevator and its counterweight as a single system driven by a motor.'' \cite{hurricane}}

In such ``piston-less'' engines the frequency of the $T_d$ in \Eq{eq:Eddington} is the inverse period of the macroscopic circulation.  As long as the external temperature gradient persists, the system will lack a mechanical equilibrium.  The resulting cyclic motion, which resembles the turning of a turbine by a steady flow, meets Andronov's definition of ``self-oscillation'', but it might be preferable to describe it as a ``self-rotation''.  Such a distinction allows us to reserve the term ``self-oscillation'' for systems, like the piston in \Fig{fig:Rayleigh}, that oscillate about a dynamically unstable equilibrium (see the discussion of limit cycles and Hopf bifurcations in \cite{SO}).

The dynamics of a thermal self-rotor depend on the macroscopic circulation, which cannot be interrupted without bringing the engine to a halt.  Since an illuminated solar cell generates a voltage in an open circuit, it evidently cannot be described as a self-rotation of this sort.  That the limit efficiency of the solar cell clearly depends on the temperature of the silicon crystal, rather than on the temperature in the external circuit, underscores the need to identify a self-oscillating piston within the solar cell.  Note also that externally reversing the circulation of a thermal self-rotor makes it into a refrigerator, which is not what happens when a solar cell is reverse-biased.

\begin{figure} [tb]
\begin{center}
	\includegraphics[width=0.25 \textwidth]{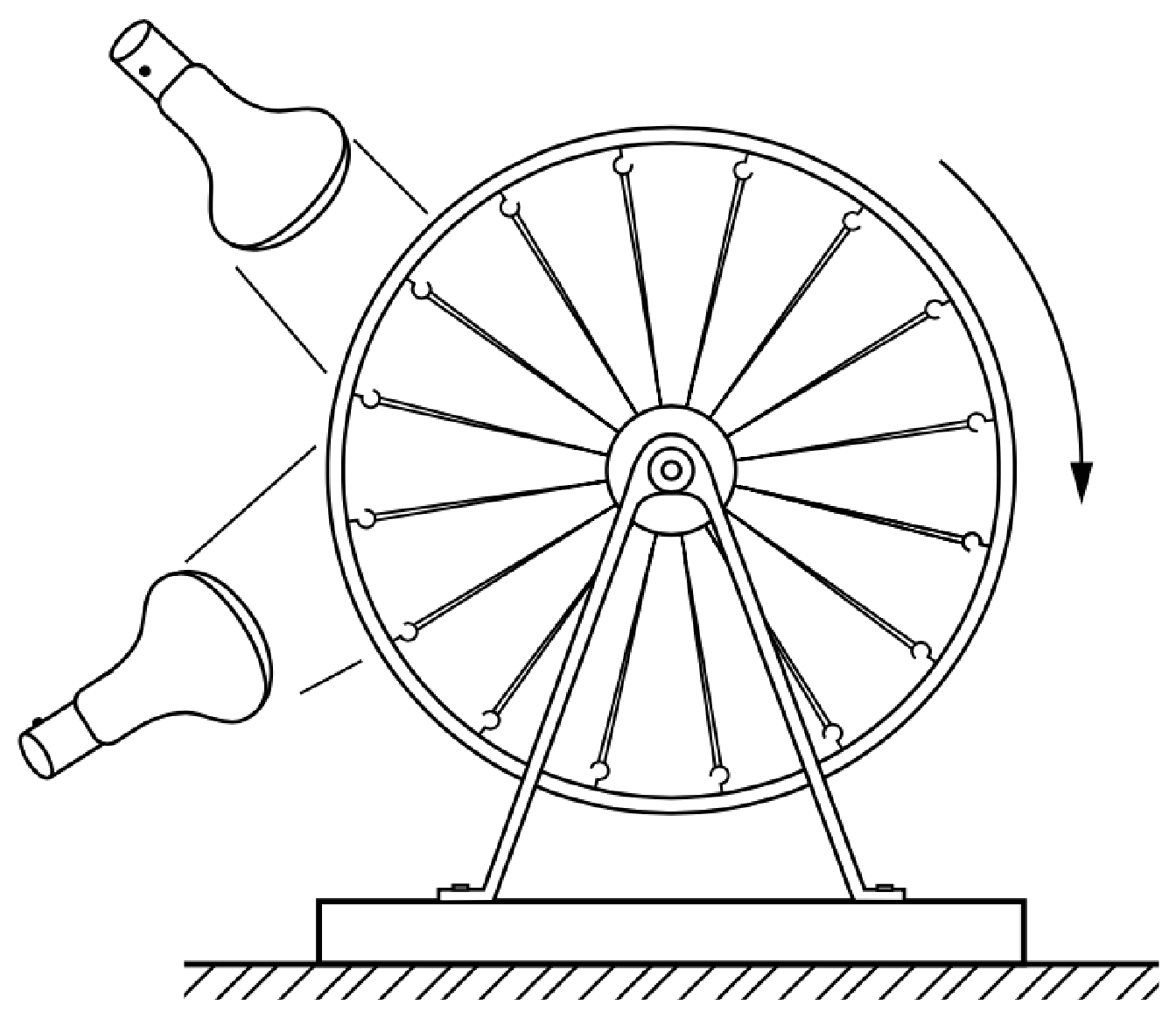}
\end{center}
\caption{\small Rubber-band heat engine as self-rotor.  Image taken from \cite{Feynman}, used here with permission.\la{fig:rubber-band}}
\end{figure}

\section{Hydrodynamical model}
\la{sec:hydrodynamical}

The textbook description of the operation of solar cells, as illustrated in \Fig{fig:textbook}, is framed in terms of the behavior of single electrons.  But one might expect the macroscopic flow of current in a circuit to result from the collective behavior of large numbers of electrons.  As we shall see, a plausible model of the self-oscillation of a piston within the solar cell also provides a hydrodynamical picture of the current pumping.

\subsection{Analogy to putt-putt pump}
\la{ssec:puttputt}

Rather than the toy boat described in \cite{Piot,FinnieCurl}, let us consider a variation of the same mechanism that can serve as a water pump.  As shown in \Fig{fig:putt-pump}, an internal tank is partly filled with water, leaving a bubble of air and steam that acts as the working substance.  The tank is connected to two pipes, one submerged in a lower reservoir, while the other is connected to an upper reservoir.  If the heat of the flame is above some minimal threshold, but not so high that all the liquid is driven out, then the level of water in the tank will self-oscillate.  (The name ``putt-putt'', or ``pop-pop'', comes from the fact the oscillation of the gas pressure may cause a noisy vibration.)  The resulting flow in the pipes is rectified by one-way valves, causing water to be pumped from the lower to the upper reservoir.

To understand phenomenologically the operation of this putt-putt pump, let us denote by $x$ the height of the liquid water in the internal tank, with $x=0$ corresponding to its equilibrium position.  Without a flame heating the water, $x$ experiences a restoring force due to the increase in the pressure of the gas when its volume decreases at constant temperature (Boyle's law).  For small oscillations we may neglect nonlinearities, giving simply
\be
\ddot x + \gamma \dot x + \omega^2 x = 0 ~, \la{eq:sho}
\ee
where $\omega$ is a resonant frequency and $\gamma$ a damping coefficient given by the friction on the moving water.  Heating the tank with the flame changes the dynamics by introducing an $x$-dependent variation in the quantity of steam in the gas bubble within the tank. \cite{FinnieCurl}

\begin{figure} [t]
\begin{center}
	\includegraphics[width=0.45 \textwidth]{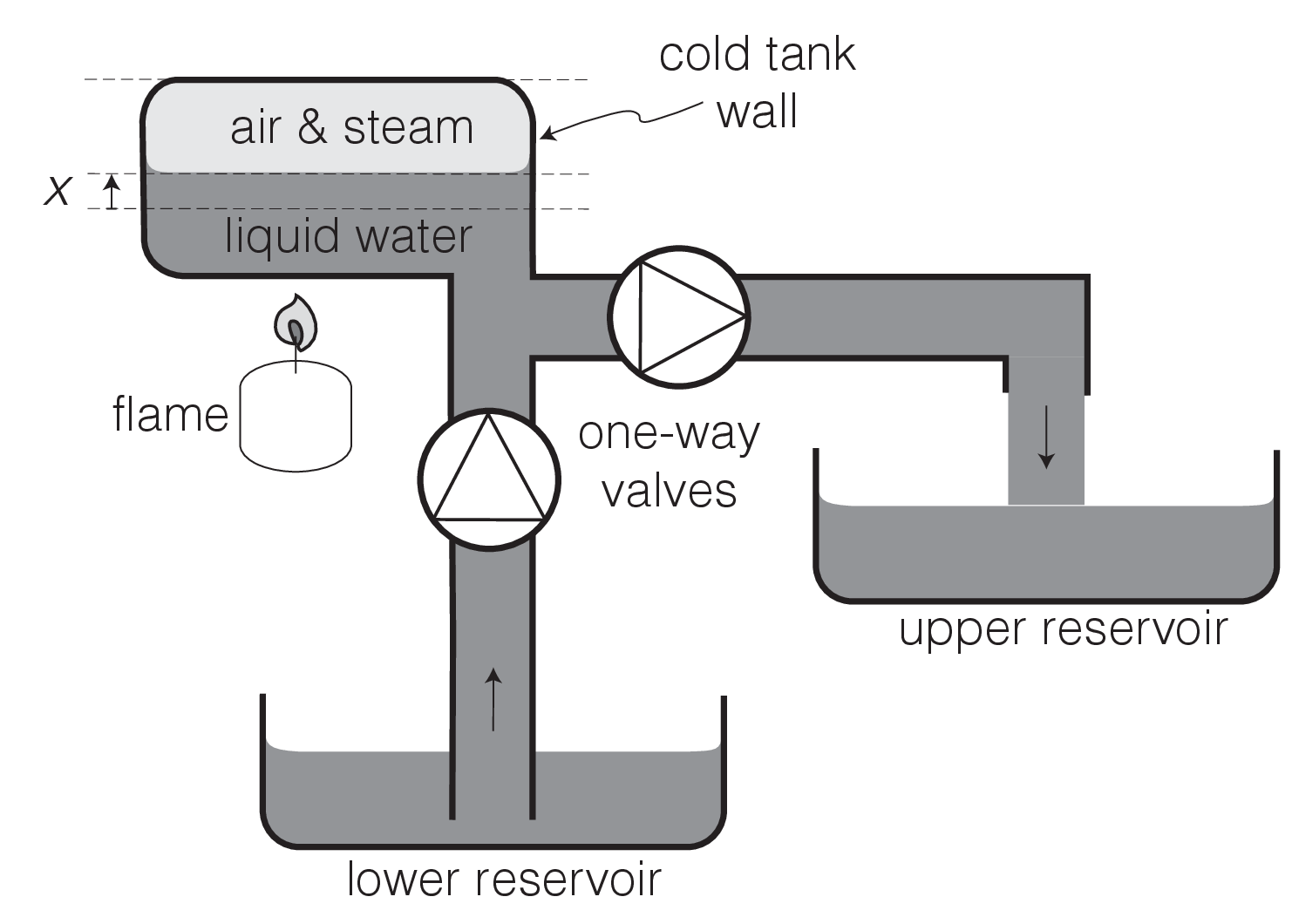}
\end{center}
\caption{\small Model of a water pump, driven by a putt-putt heat engine, which can raise water from a lower to an upper reservoir.\la{fig:putt-pump}}
\end{figure}

Let $N_0$ be the quantity (i.e., number of moles) of steam when the liquid water is at its equilibrium level ($x=0$).  Most of the steam remains in thermal equilibrium with the hot water below, and therefore at a fixed temperature.  Under the action of the flame, small oscillations of $x$ therefore obey
\be
\ddot x + \gamma \dot x + \omega^2 x =  A \left( N_0 - N \right) ~,\la{eq:x} 
\ee
where $A$ positive and constant, while $N$ is the instantaneous quantity of steam.  The variation in time of the quantity of steam may be expressed as
\be
\dot N  =  - \Gamma(x) {N} + B(x) ~, \la{eq:N}
\ee
where the rate of condensation is denoted by $\Gamma (x) \geq 0$ and the rate of evaporation by $B(x) \geq 0$.  Equations \eqref{eq:x} and \eqref{eq:N} jointly describe the dynamics of $x$.  As we shall see in \Sec{ssec:stability}, for appropriate values of the parameters of this model, the heat from the flame can sustain the self-oscillation of $x$ about its equilibrium.

A closely analogous hydrodynamical model can be applied to the solar cell.  The collective oscillation of a relatively dense electron gas (or rather quantum fluid) in the $n$-type layer, in contact with a dilute electron gas in the $p$-type layer, can be described by a single degree of freedom: the position $x$ of the depletion layer (which we simply call the interface).  The resonant $\omega$ corresponds to the plasma frequency 
\be
\omega_{\rm p} = \sqrt{\frac{n_e e^2}{m_\ast \epsilon}} ~, 
\la{eq:plasma_f}
\ee
where $n_e$ is a density of electrons, $m_\ast$ their effective mass, and $\epsilon$ is the material's dielectric constant.  Several experimental groups have observed such plasma oscillations at $p$-$n$ junctions, with frequencies in the THz domain, by hitting the junction with an optical laser pulse and observing the coherent electromagnetic radiation emitted as the plasma oscillation subsequently rings down \cite{plasmaoscillations1, plasmaoscillations2}.  It has also been argued that a reverse-biased $p$-$n$ junction with current injection may self-oscillate with a frequency of that order \cite{SO-plasma}.  However, self-oscillation in an illuminated solar cell ---the key prediction of our model--- has not been reported.  Note that the shorter scale of spatial coherence for the fluid's motion and the far higher frequency make such plasma oscillation in an illuminated cell considerably more difficult to study than the vibration of the liquid-steam interface in the putt-putt engine, even though the two phenomena are otherwise similar in nature.

In an illuminated solar cell, the $p$-type layer acts as a photon absorber.  The dynamics of the oscillation in $x$ can be described by Eqs.\ \eqref{eq:x} and \eqref{eq:N}, with $N$ now corresponding to the number of photo-generated electrons in the absorber and $N = N_0$ at equilibrium.  The oscillating bulk of the electronic gas acts like an electron pump, much like the oscillating water level in the tank can drive water in the pipes in \Fig{fig:putt-pump}.  In the case of the water pump, the flow of water is made unidirectional by the action of the valves.  In the case of the solar cell, the junctions in the front and back of the absorber act as diodes, rectifying the oscillatory current, as will be further discussed in \Sec{ssec:rectification}.

Much like the pump in \Fig{fig:putt-pump} moves water against gravity ---and can therefore maintain a cyclic circulation if the reservoirs are connected--- the piston of the solar cell drives electrons against the electrostatic field between opposite terminals of the illuminated solar cell, thereby powering the DC in the closed circuit.  Note that while the putt-putt's working substance is the steam in the tank, the fluid being pumped is the liquid water.  In our model of the solar cell, the same electron gas that acts as the working substance is being pumped from one terminal of the cell to the other.  Thus, a more perfect analogy of our model of the solar cell would be a putt-putt engine that pumped steam rather than liquid water, but we have preferred to stick with the familiar water-pumping device in our discussion.

\subsection{Stability analysis}
\la{ssec:stability}

To find the conditions under which $x$ will self-oscillate, we linearize Eqs.\ \eqref{eq:x} and \eqref{eq:N}, putting
\be
\Gamma(x) = \Gamma  + g x , \quad B(x) = B + b x , \quad n \equiv N - N_0 ~,
\la{eq:lin}
\ee
with $N_0 = B / \Gamma$.  For convenience, we also pick units of time and position such that $\omega =1 $ and $A = 1$.  Equations \eqref{eq:x} and \eqref{eq:N} are then replaced by a set of three linear, first-order differential equations:
\bea
\dot x &=& v \la{eq:xx} \\
\dot v &=& -x - \gamma v - n \la{eq:vv} \\
\dot n &=& f x - \Gamma n ~, \la{eq:nn}
\eea
with feedback parameter
\be
f \equiv b - g N_0 ~.
\la{eq:b}
\ee

The stability of this system is determined by the eigenvalues $\lambda$ of the corresponding $3 \times 3$ matrix:
\be
\bigl[\lambda(\lambda + \gamma) +1\bigr] (\lambda + \Gamma) + f = 0 ~.
\la{eq:poly}
\ee
When the real part of an eigenvalue $\lambda$ is positive, the equilibrium $x=0$ is unstable and the amplitude of small perturbations grows exponentially with time.  If that eigenvalue has non-zero imaginary part, this corresponds to a self-oscillation, whose steady amplitude is determined by the non-linearities in Eqs.\ \eqref{eq:x} and \eqref{eq:N}. \cite{SO}

\begin{figure} [t]
\begin{center}
	\subfigure[]{\includegraphics[width=0.42 \textwidth]{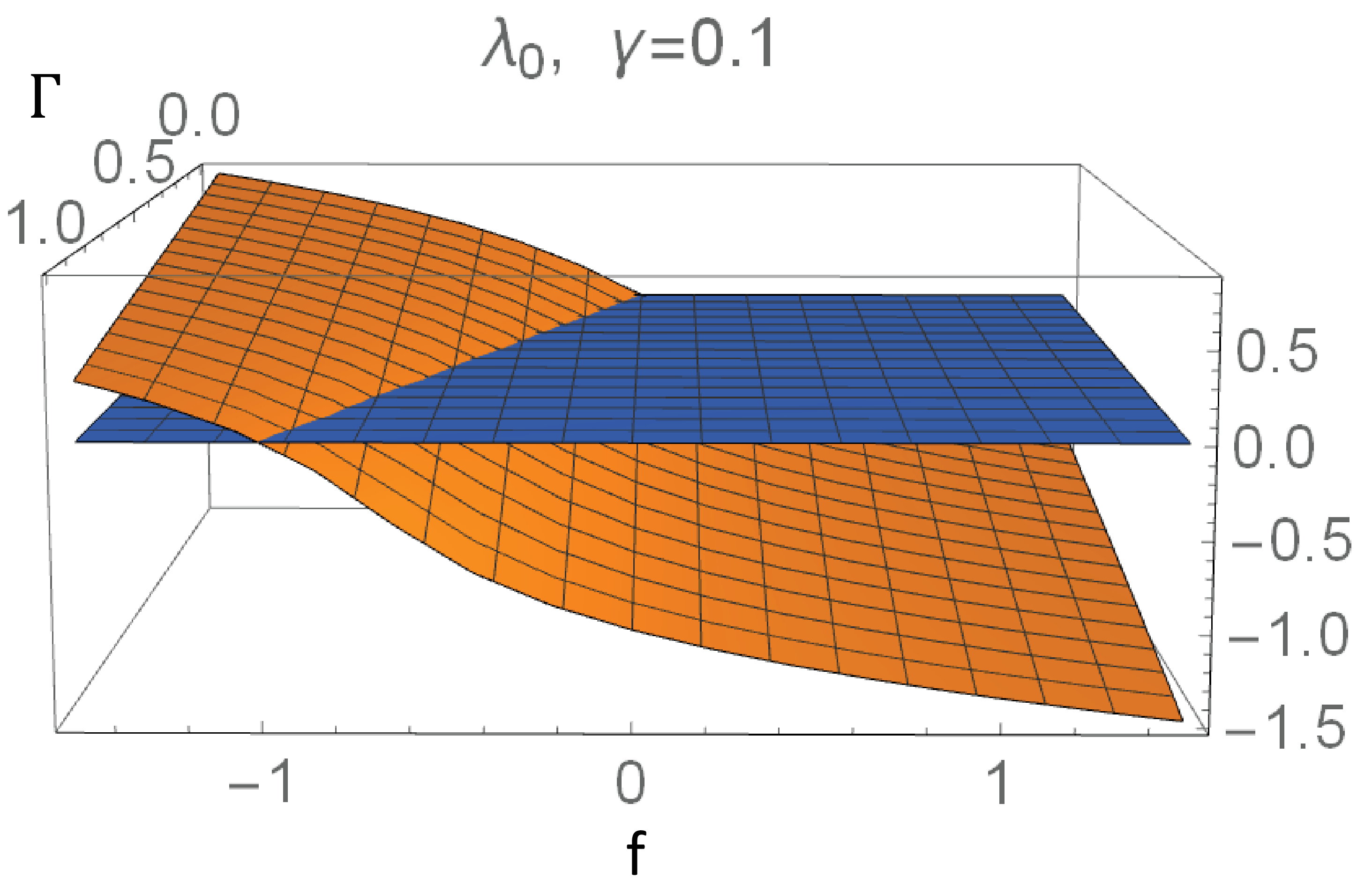}} \hskip 1 cm
	\subfigure[]{\includegraphics[width=0.42 \textwidth]{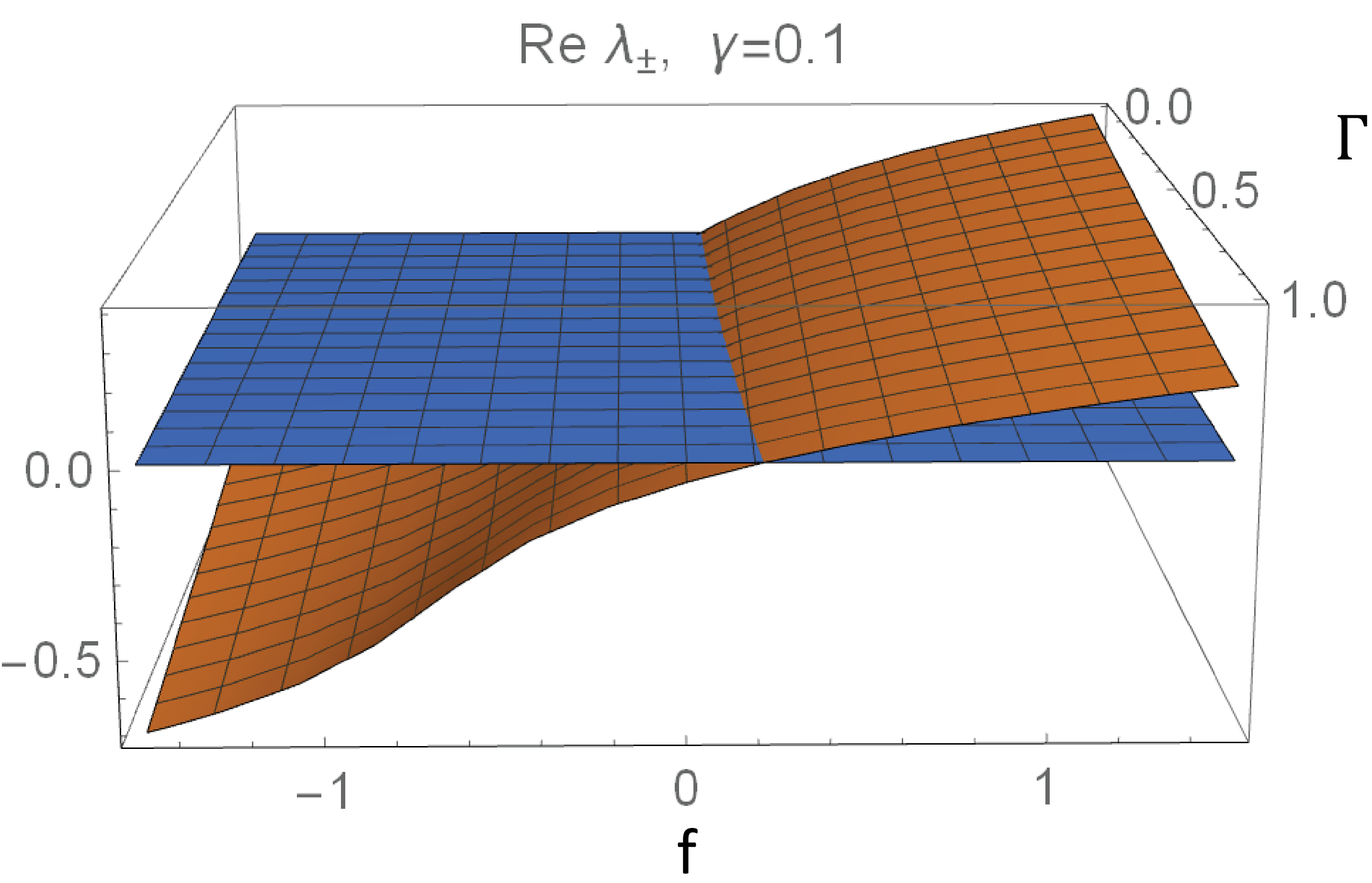}}
\end{center}
\caption{\small The three solutions of \Eq{eq:poly}: (a) $\lambda_0$ as a function of the parameters $f$ and $\Gamma$, for fixed $\gamma = 0.1$.  (b) $\Re \lambda_\pm$ as a function of $f$ and $\Gamma$, for $\gamma = 0.1$.  The zero plane is shown in blue.\la{fig:eigenvalues}}
\end{figure}

Approximate solutions of \Eq{eq:poly} can be found analytically for $\Gamma, \gamma , |f| \ll 1$, in which case there will be one real eigenvalue $\lambda_0$ and two complex, conjugate eigenvalues $\lambda_{\pm}$ close to $\pm i$. Then, up to higher order corrections,
\be
\lambda_0 \simeq - \left( \Gamma + f \right)
\la{eq:eigen0}
\ee
and 
\be
\lambda_{\pm} \simeq \pm i +\frac{1}{2} \left( f - \gamma \right) ~.
\la{eq:eigen}
\ee
Self-oscillation therefore occurs for a positive feedback parameter
\be
f > \gamma ~.
\la{eq:self_cond}
\ee

For a $\Gamma, \gamma,$ and $|f|$ not small, the eigenvalues may be computed numerically.  The values of $\lambda_0$ and $\Re \lambda_\pm$ as functions of $f$ and $\Gamma$ are shown graphically in \Fig{fig:eigenvalues}, for fixed $\gamma = 0.1$.  When $\lambda_0 < 0$ and $\Re \lambda_\pm > 0$ the system self-oscillates about $x = 0$.  In physical units, \Eq{eq:self_cond} corresponds to
\be
f = \frac{A \left( b - g N_0 \right)}{\omega^2} >  \gamma ~.	
\la{eq:self_cond1}
\ee

\subsection{Feedback mechanism}
\la{ssec:feedback}

Let us now consider the physics of the positive feedback responsible for the self-oscillation of the water pump and the solar cell.  In both cases, the variation of $x$ modulates the heat flow in a such a way that a net mechanical work $W > 0$ can be extracted during a full cycle, in accordance with the Rayleigh-Eddington criterion discussed in \Sec{ssec:Rayleigh}.

\subsubsection{Putt-putt pump}
\la{sssec:putt}

As originally proposed in \cite{FinnieCurl}, a fall in the water level $x$ inside the putt-putt engine increases the rate of condensation of steam by increasing the surface of contact between the steam and the wall of the tank, which in turn is in contact with the colder environment.  Thus, the condensation rate $\Gamma (x)$ grows when $x$ decreases, corresponding to $g < 0$ in \Eq{eq:lin}.  The rate of evaporation is independent of the water level, so that $b = 0$ in \Eq{eq:lin}.  Therefore the putt-putt engine exhibits a positive feedback ($f > 0$) and will self-oscillate as long as this exceeds the damping of the oscillation of the water level ($f > \gamma$).

As underlined in \cite{SO}, the putt-putt engine optimizes the Rayleigh-Eddington criterion, because the rate of rejection of heat (effected by condensation) is greatest when the working gas reaches its maximum volume.  Note that when the volume of the working gas increases, its average temperature decreases, even though the expansion is not adiabatic, because more of the gas comes into contact with the cold tank wall.

\subsubsection{Solar cell}
\la{sssec:solar}

For the operation of the solar cell as a self-oscillating heat engine to optimize the Rayleigh-Eddington criterion, the rate at which conducting pairs are created must increase when $x$ increases, i.e., when the interface moves into the absorber.  When the collective oscillatory motion of the electrons in the $n$ phase (i.e., the piston $x$) compresses the absorber ($p$) phase, some of the photo-generated electrons will ``roll'' down the potential towards the negatively charged terminal, as shown in \Fig{fig:feedback}.  At the same time, holes in the absorber are pushed towards the positively charged terminal, thus preserving the local charge balance.  This drives current out of the terminals and into the external circuit, as in the textbook account of the photovoltaic effect.  Thus, as the piston compresses the absorber the number of holes in its valence band is reduced.  This increases the rate of photo-generation of conducting pairs (i.e., of heat absorption by the working substance), corresponding to $b > 0$ in \Eq{eq:lin}.  The narrowing of the band gap (see \cite{Sze}) when the absorber's temperature and density increase under compression could, at least in some circumstances, play an accessory role in the positive feedback.

\begin{figure} [t]
\begin{center}
	\includegraphics[width=0.55 \textwidth]{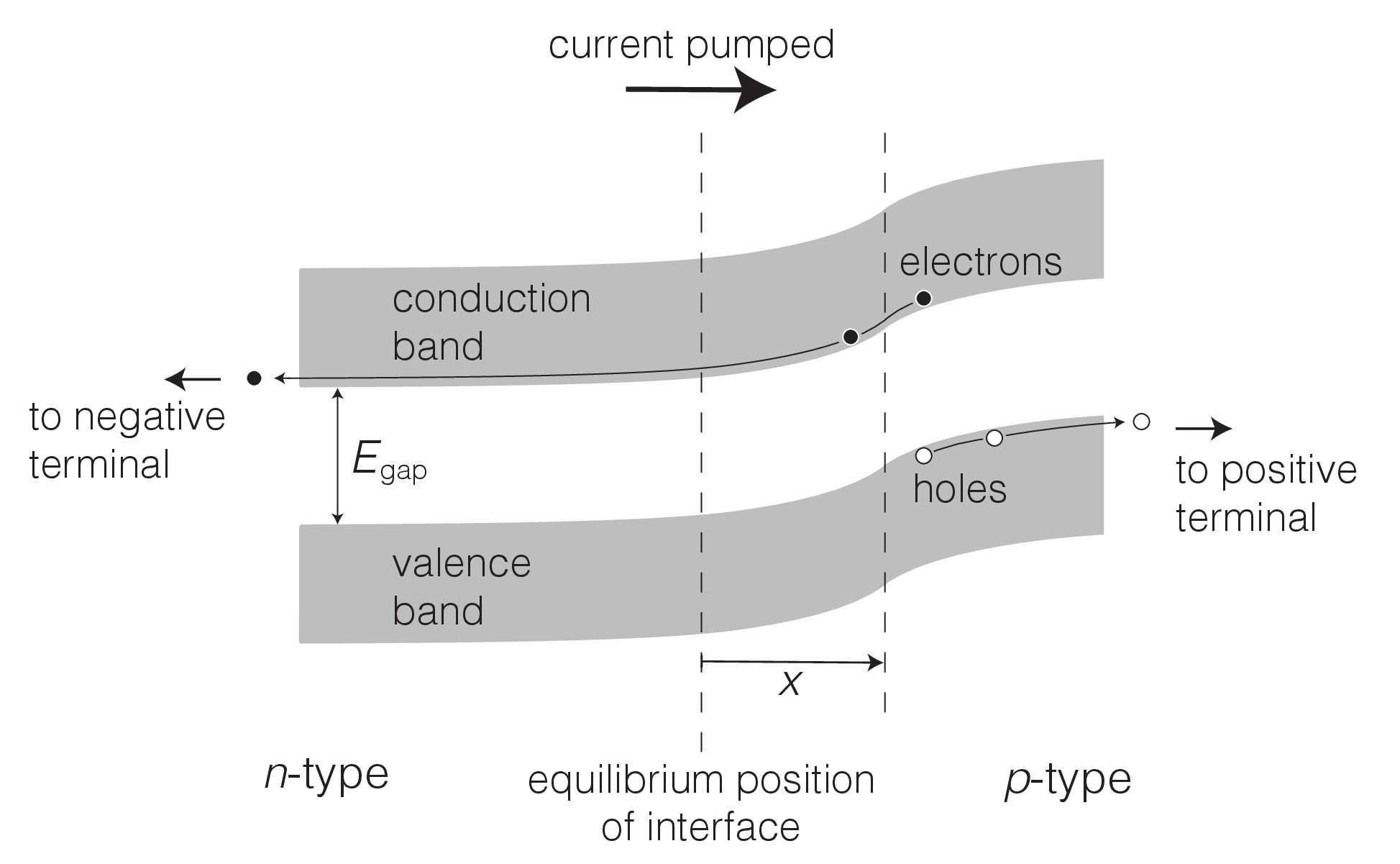}
\end{center}
\caption{\small The displacement of the solar cell's $p$-$n$ interface away from its equilibrium position ($x=0$) in a direction that compresses the $p$-type phase ($x > 0$) pumps current in the direction indicated, while reducing the number of valence holes in the absorber.\la{fig:feedback}}
\end{figure}

\subsection{Rectification}
\la{ssec:rectification}

The putt-putt toy boat, though valveless, moves forward because it gains net {\it momentum} only during the phase in which water is expelled \cite{sprinkler}.  In the putt-putt pump of \Fig{fig:putt-pump} it was necessary to introduce one-way valves in order to transform the oscillation of the water level into a directional flow of water from the lower to the upper reservoir.  A single valve suffices, though the efficiency of the pump is increased by using two valves.  At extremely high frequencies of the putt-putt pump's oscillation, the resulting flow might not pulsate and only a steady difference in the respective hydrodynamical pressures of the inflow and the outflow might be observed.

The mechanism of rectification in our model of the photovoltaic effect is closely analogous to this.  The interface between the layers of the cell acts as a diode, making it easier for positive current to flow from $n$ to $p$.  When the piston compresses the absorber, current is pushed out of the terminals, as described in \Sec{sssec:solar}.  When, in the next phase of the cycle, the piston allows the absorber to expand, the current is not reversed because conducting electrons will not easily move up the slope of the potential at the interface.  Note that modern solar cells often add a highly doped $p$-type layer on the other side of the absorber, which acts as a second diode \cite{Sze}.  This improves the efficiency by minimizing the reversal of current during the cycle.

\section{Thermodynamic efficiency}
\la{sec:efficiency}

The open-circuit voltage $V_{\rm oc}$ and the maximum efficiency $\eta_{\rm max}$ cannot be calculated in our simplified mechanical model, but the laws of thermodynamics impose bounds on those quantities, in terms of the band gap $\Eg$, the ambient temperature $T_1$, and the effective ``light temperature'' $T_2$.  This is one of the advantages of framing the operation of the solar cell as a thermodynamic cycle.

\subsection{Significance of gap}
\la{ssec:gap}

The semiconductor band gap is essential to the photovoltaic effect, as it introduces a time-scale separation between the fast intraband transitions (with relaxation times typically in the sub-picosecond scale) and the comparatively slow interband transitions (with relaxation times typically in the nanosecond scale).  This prevents the electron gas in the $p$-phase of the illuminated solar cell (i.e., the working substance in our heat-engine description) from thermalizing.  Only the slow transitions across the band gap are relevant for generating work.  For most purposes, the solar cell may therefore be regarded as an engine with a two-level quantum system as working substance; see \cite{Markovian1}.

Analogously, in the putt-putt engine the latent heat of evaporation introduces a time-scale separation between the fast thermalization of the gas molecules and the slow interconversion between liquid water and steam.  This prevents the steam from simply thermalizing with the hot water below.  Note that water's latent heat of evaporation is \hbox{2.2 MJ/kg}, or 0.42 eV per molecule, which is of the same order of magnitude as the band gap of a silicon solar cell, \hbox{$\Eg \simeq 1.1$ eV}.  This helps explain why, despite the solar-cell engine operating at THz frequencies and the putt-putt engine at a few Hz, the two devices generate comparable power when coupled to comparable heat baths.

\subsection{Local temperature of non-equilibrium bath}
\la{ssec:local}

The sunlight that reaches the Earth's surface constitutes a highly non-equilibrium yet stationary bath, characterized by photon population as function of frequency, $n(\omega)$.  A general quantum theory of non-equilibrium, stationary baths has recently been developed in \cite{localT}, based on the notion of a ``local'' (i.e., frequency dependent) temperature $T_{\rm loc}(\omega)$.  This local temperature is measured by a quantum thermometer with a single working frequency $\omega$, such as a two-level system or a harmonic oscillator.  When coupled to the bath characterized by $n(\omega)$, the thermometer approaches a Gibbs state with $T = T_{\rm loc}(\omega)$ given by the relation
\be
\frac{n(\omega)}{n(\omega) +1}  = e^{ - \hbar \omega/ k_B T_{\rm loc}(\omega)}~.
\la{eq:localT} 
\ee
This notion of local temperature can be used to formulate the second law of thermodynamics for open quantum systems coupled to non-equilibrium, stationary baths.  This yields an efficiency bound for heat engines operating between two baths that is completely analogous to the ordinary Carnot bound, thus providing a consistent thermodynamic interpretation of the local temperature. \cite{localT}

For sunlight at the Earth's surface we may approximate
\be
n(\omega) = \frac{\lambda}{e^{\hbar\omega/k_B T_\odot} - 1} ~,
\la{eq:roughT} 
\ee
where $T_\odot \simeq 6{,}000 ~\hbox{K}$ is the equilibrium temperature of the Sun's surface,  $\lambda = (R_\odot / R_0)^2 \simeq 2 \times 10^{-5}$ is the geometric factor giving the dilution of the photon density as solar radiation travels from its source, on a spherical shell of radius $R_\odot$, to the Earth, at a distance $R_0$ from the Sun's center.

The non-equilibrium photon bath described by \Eq{eq:roughT} was described in \cite{LandsbergTonge} as ``diluted black-body radiation".  In that work, an ``absolute temperature'', coinciding with the local temperature of \Eq{eq:localT}, was derived using different thermodynamical arguments from those of \cite{localT}.  Figure \ref{fig:effective} gives a plot of the local temperature corresponding to \Eq{eq:roughT}.

\begin{figure} [t]
	\begin{center}
		\includegraphics[width=0.45 \textwidth]{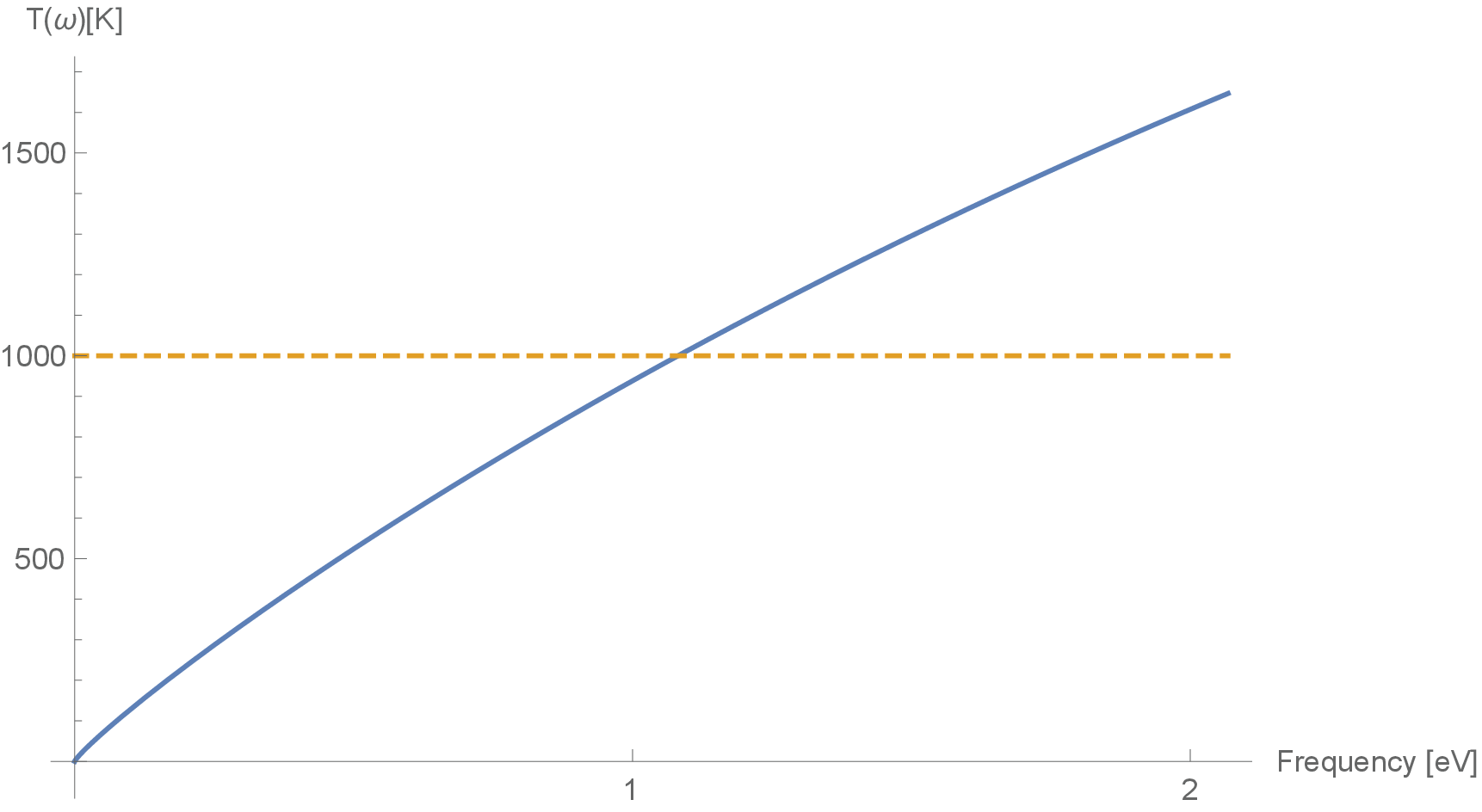}
	\end{center}
\caption{\small Plot of the local temperature $T_{\rm loc}$ as function of photon frequency $\omega$, according to \Eq{eq:roughT}, for solar radiation reaching the Earth's surface.  The dashed line indicates the local temperature evaluated at the silicon band gap, $\omega = 1.1$ eV.\la{fig:effective}}
\end{figure}

\subsection{Ultimate efficiency}
\la{ssec:ultimate}

The efficiency of the solar cell is bounded not only by the thermodynamic Carnot factor $\eta_{\rm th}$, but also by an ``ultimate efficiency'' factor $\eta_{\rm u}$ that reflects the fact that only a transition corresponding to $\Eg$ can be used to generate work.  A solar photon with energy $E < \Eg$ is not absorbed, while absorption of a photon with $E > \Eg$ is followed by the rapid dissipation of the excess $E - \Eg$ into the cold phonon bath.\footnote{Depending on the details of the design of the solar cell, it may be possible to avoid dissipation of some of the excess energy in photons with $E > \Eg$; see, e.g., \cite{Abrams}.  We will not consider that possibility here.}  The factor $\eta_{\rm u}$ is equal to $\Eg$ times the number of photons absorbed by the solar cell, divided by the total energy of the incident photons.  For a silicon solar cell under standard illumination conditions $\eta_{\rm u} \simeq 44\%$.  \cite{SQ}

It might be interesting to investigate whether the putt-putt engine has an $\eta_{\rm u}$ factor analogous to the solar cell's.  The latent heat of evaporation has been plausibly explained as resulting from the change in the potential energy of the liquid's surface as an additional molecule is pulled into the liquid bulk \cite{Garai}.  It may be, therefore, that the pressure exerted on the liquid by the putt-putt's working gas results, not from elastic collisions (as it would if the gas pushed on a solid piston), but rather from surface tension stretching flat the liquid's surface, after a molecule originating in the gas comes into contact with it.  In that process, the molecule's kinetic energy would be dissipated.  This process would consume gas, much like photo-generated conducting pairs recombine due to an external load in the circuit connected to an illuminated solar cell.

\subsection{Carnot factor}
\la{ssec:Carnot}

\begin{figure} [t]
\begin{center}
	\includegraphics[width=0.3 \textwidth]{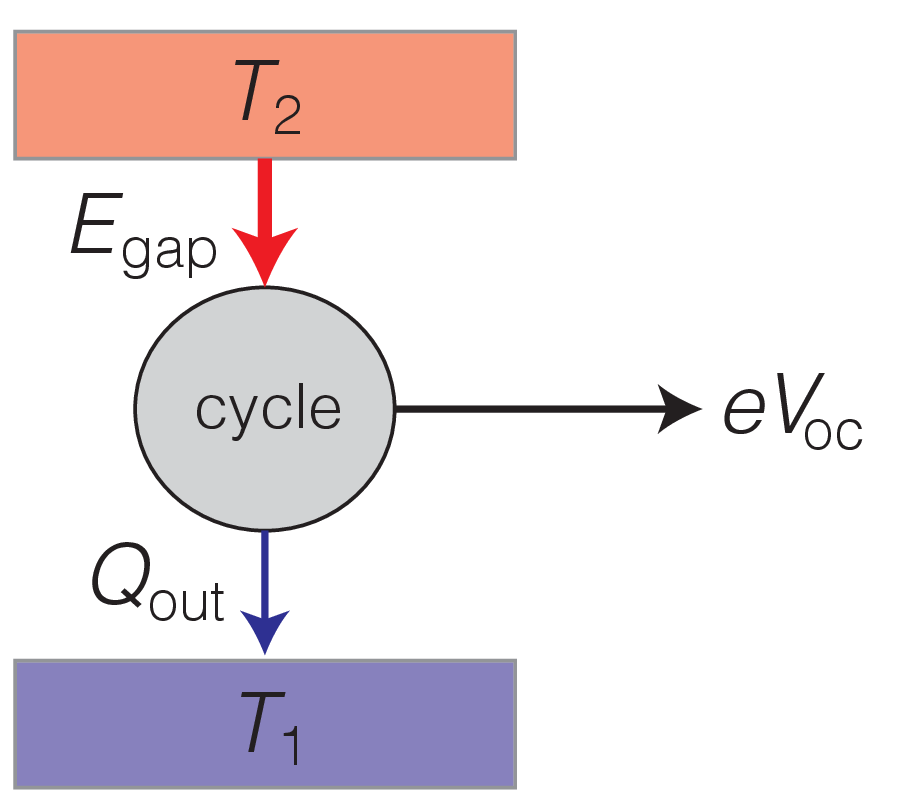}
\end{center}
\caption{\small Photo-generation of a conducting pair in an illuminated solar cell is accompanied by absorption of an amount $\Eg$ of potentially usefully heat from sunlight, at a high effective temperature $T_2$.  For a cyclic process, the laws of thermodynamics require that this be accompanied by rejection of heat $Q_{\rm out}$ into the environment at room temperature $T_1$, with $Q_{\rm out} / T_1 \geq \Eg / T_2$.  The difference $\Eg - Q_{\rm out}$ is available to perform a maximum amount of electrical work per conducting pair equal to $eV_{\rm oc}$, where $V_{\rm oc}$ is the open-circuit voltage of the illuminated cell.\la{fig:Carnot-cell}}
\end{figure}

The quantity $e V_{\rm oc}$ is the maximum work that may be obtained per electron leaving the bottom of the absorber's conduction band and returning to the absorber by recombination with a hole at the top of the valence band \cite{Fermi}.  As explained in Secs.\ \ref{ssec:gap} and \ref{ssec:ultimate}, only $\Eg$ per electron will be available for generating work, because a photon with energy below $\Eg$ is not absorbed, whereas the excess $E - \Eg$ in a photon with energy $E > \Eg$ is very quickly dissipated.  Combining this with the fundamental thermodynamic bound on the cyclic conversion of heat into work gives
\be
eV_{\rm oc} \leq \Eg \cdot \eta_{\rm th} ~, 
\la{eq:Voc}
\ee
where
\be
\eta_{\rm th} = 1 - \frac{T_1}{T_2}
\la{eq:Carnot-general}
\ee
is the familiar Carnot factor with ambient temperature $T_1$ and effective light temperature $T_2$.  The corresponding thermodynamic diagram is shown in \Fig{fig:Carnot-cell}.

Due to its thermodynamical features (see \Sec{ssec:local}) we can identify the effective light temperature $T_2$ with the local temperature of \Eq{eq:roughT}, evaluated at the frequency $\omega_0 = \Eg / \hbar$ for $\Eg \simeq 1$ eV:
\be
T_2 \simeq T_{\rm loc}(\omega_0) \simeq 1{,}000 ~\hbox{K} ~.
\la{eq:T1}
\ee
The Carnot factor then comes out to
\be
\eta_{\rm th} \simeq \left( 1 - \frac{300 ~\hbox{K}}{1{,}000 ~\hbox{K}} \right) = 70\% ~.
\la{eq:Carnot-solar}
\ee
Combining this with the ultimate efficiency factor $\eta_{\rm u}$ of \Sec{ssec:gap} gives a bound on the overall maximum efficiency of
\be
\eta_{\rm max} = \eta_{\rm u} \cdot \eta_{\rm th} \simeq 31\% ~,
\la{eq:etamax}
\ee
in agreement with the detailed balance bound of Shockley and Queisser. \cite{SQ}

\subsection{Dissipative losses}
\la{ssec:losses}

Dissipation within the cell implies generation of entropy $\sigma > 0$ in \Eq{eq:Clausius} and therefore reduces the cell's efficiency relative to the upper limit set by the Rayleigh-Eddington criterion of \Eq{eq:Eddington}.  It is always possible, in principle, to improve the cell's efficiency by reducing such dissipative losses.

We have already pointed out one form of dissipation present in the illuminated solar cell: process {\bf 2} in \Fig{fig:textbook}, which contributes to the ultimate efficiency factor discussed in \Sec{ssec:ultimate}.  Meanwhile, friction on the plasma oscillation is included in the damping factor $\gamma$ of \Eq{eq:sho}.  In both cases, energy is dissipated into the bath of phonons at room temperature.  The pair recombination described by the function $\Gamma(x)$ in \Eq{eq:N} is also associated with dissipation into the phonon bath, but some of the energy released by that recombination may be radiated as coherent sub-millimeter electromagnetic radiation (the analog of the sound generated by the putt-putt engine, giving that engine its same), which should be counted as part of the cell's work output.

In this work, which was focused on general thermodynamic considerations, we did not attempt to calculate the solar cell's internal resistance.  Including this in our calculation of the efficiency would require examining the detailed physical interactions, leading to a determination of the current-voltage ($I$-$V$) characteristic.  This is left for future work.  Note that the electrical work outputted by the cell can be stored (e.g., by charging a battery), transformed into mechanical energy (e.g., by running a DC motor) or dissipated by the resistance of the load attached to the cell's terminals.  The details of such processes are independent of the solar cell's operation as a heat engine.

\section{Discussion}
\la{sec:discussion}

We proposed a simple model of the photovoltaic effect, consistent with the laws of thermodynamics as well as with a hydrodynamical description of the macroscopic current generated by that effect.  For this it was necessary to introduce a new element into the dynamics of a solar cell: a self-oscillating piston in the picture of the solar cell as a cyclic heat engine.  This piston is the interface between the $n$-type and $p$-type phases of the semiconductor, an interface that has been experimentally observed to oscillate with a resonant frequency in the order of 1 THz.

We have shown, in a dynamically complete model, that this piston may self-oscillate when the solar cell is exposed to heat in the form of light at an effective temperature significantly above the cell's ambient temperature.  The piston gains mechanical energy after a full period of its oscillation because it modulates the rate at which the $p$-type phase absorbs heat from incident light.  The directionality of the electric potential in the vicinity of that interface causes that oscillation to pump current preferentially in one direction, allowing it to drive a DC along a closed circuit connected to the illuminated cell.

This picture is consistent with the basic features of the photovoltaic effect.  It also allows for a simple thermodynamic derivation of a limiting efficiency, consistent with that obtained by Shockley and Queisser by a more complicated argument of detailed balance.  We have further argued for the plausibility of this model by pointing to its close analogy to the operation of the familiar putt-putt engine, in which a liquid-steam interface serves as the self-oscillating piston.

An obvious consequence of our model is that the operation of an illuminated solar cell should be accompanied by a weak THz (i.e., sub-millimeter) radiation with an amplitude squared roughly proportional to the cell's power output.  Moreover, it should also be possible to increase a cell's power output by externally driving that oscillation at its resonant frequency.  If the power in the THz frequency range is not entirely dissipated by the hydrodynamic response of the electronic fluid, it is conceivable that small residual oscillations with that high frequency might be seen in the cell's output voltage.  Any of these effects is likely to be difficult to measure over the various kinds of noise that accompany the operation of an illuminated solar cell, but we hope to have offered compelling physical arguments why experimentalists may wish to look for them.

At this stage we are not in a position to make concrete suggestions for improving the performance of solar cells.  However, if we are correct in identifying the plasma oscillation at the $p$-$n$ junction of the semiconductor as an essential component in the operation of the cell as a heat engine, taking its dynamics into consideration may facilitate future modifications and improvements to existing solar energy technologies. \\

{\bf Acknowledgements:}  We thank Marek Grinberg, Iwo Bia{\l}ynicki-Birula, Daniel Vanmaekelbergh, Gavin Conibeer, Klaus Lips, Robert Jaffe, Pavel Chvykov, and Esteban Avenda\~no for discussions, as well as Michael Gottlieb (editor of the online {\it Feynman Lectures on Physics}) for permission to use \Fig{fig:rubber-band}.  DG-K's work was supported in part by the Center for Excitonics, an Energy Frontier Research Center funded by the US Department of Energy under award DE-SC0001088 (Solar energy conversion process), and by CONACYT (Quantum thermodynamics).  AJ gratefully acknowledges the hospitality of MIT's Center for Theoretical Physics while some of this work was completed.


\bibliographystyle{aipprocl}   

\end{document}